\documentclass[preprint, superscriptaddress, showpacs,preprintnumbers,amsmath,amssymb,prd]{revtex4}
\usepackage{graphicx}

\begin{document}

\thispagestyle{empty}

\title{Constraining axion coupling constants
from measuring the Casimir interaction between polarized
test bodies
}

\author{V.~B.~Bezerra}
\affiliation{Department of Physics, Federal University of Para\'{\i}ba,
C.P.5008, CEP 58059--970, Jo\~{a}o Pessoa, Pb-Brazil}
\author{
G.~L.~Klimchitskaya}

\affiliation{Central Astronomical Observatory
at Pulkovo of the Russian Academy of Sciences,
St.Petersburg, 196140, Russia}
\affiliation{Institute of Physics, Nanotechnology and
Telecommunications, Peter the Great St.Petersburg
Polytechnic University, St.Petersburg, 195251, Russia}
\affiliation{Department of Physics, Federal University of Para\'{\i}ba,
C.P.5008, CEP 58059--970, Jo\~{a}o Pessoa, Pb-Brazil}

\author{
 V.~M.~Mostepanenko}
\affiliation{Central Astronomical Observatory
at Pulkovo of the Russian Academy of Sciences,
St.Petersburg, 196140, Russia}
\affiliation{Institute of Physics, Nanotechnology and
Telecommunications, Peter the Great St.Petersburg
Polytechnic University, St.Petersburg, 195251, Russia}
\affiliation{Department of Physics, Federal University of Para\'{\i}ba,
C.P.5008, CEP 58059--970, Jo\~{a}o Pessoa, Pb-Brazil}

\author{C.~Romero}
\affiliation{Department of Physics, Federal University of Para\'{\i}ba,
C.P.5008, CEP 58059--970, Jo\~{a}o Pessoa, Pb-Brazil}

\begin{abstract}
We propose an experiment for measuring the effective Casimir
pressure between two parallel SiC plates with aligned nuclear
spins. The prospective constraints on an axion-neutron
coupling constant for both hadronic and GUT axions are
calculated using the process of one-axion exchange. For this
purpose, a general expression for the additional pressure
arising between two polarized plates due to the exchange of
one axion between their constituent fermions is derived. We
demonstrate that only the polarization component perpendicular
to the plates contribute to the pressure. The obtained pressure
can be both repulsive and attractive depending on whether the
polarizations of both plates are unidirectional or directed in
opposite directions. It is shown that although the
constraints on an axion-electron coupling obtained in the case
of magnetized plates are not competitive, the constraints on an
axion-neutron coupling found for plates with polarized nuclear
spins are of the same order of magnitude of those obtained
previously for the GUT axions alone using the process of
two-axion exchange. The proposed experiment allows also to
strengthen the presently known constraints on the axion-neutron
coupling constants of GUT axions by using both processes
of one- and two-axion exchange.
\end{abstract}

\maketitle
\section{Introduction}

Axions are light pseudo-scalar particles which were predicted long
ago \cite{1,2}, but up to the present have not been found
experimentally,
in spite of repeated attempts. Much attention given to axions on
both theoretical and experimental sides is determined by the
important role  they play in elementary particle physics,
gravitation and cosmology. In QCD, axions explain the absence of
strong CP violation and of the large electric dipole moment of the
neutron \cite{3,4,5}. They also provide the most natural solution
for the problem of dark matter \cite{3,6}.

The originally introduced QCD axions \cite{1,2} are
pseudo-Nambu-Goldstone bosons. They interact with fermions via
a pseudo-vector Lagrangian. Rather quickly, the prediction of
Refs.~\cite{1,2} was constrained to a very narrow band in
parameter space \cite{7},
and was generalized to  ``invisible"
hadronic axions \cite{8,9} possessing much smaller interaction
constants.
Another type  of the so-called grand unified theory (GUT) axions, or
axion-like particles, was proposed in Refs.~\cite{10,11}.
In the context of many models,
the GUT axions interact with fermions via a pseudo-scalar
Lagrangian \cite{3}.

Experimental searches for axions and axion-like particles are
numerous and diverse \cite{4,5,7,12}. They are based on the
interaction of axions with photons, electrons and nucleons.
Many experiments use the so-called helioscopes and haloscopes
\cite{13}, designed to register axions generated in the Sun
\cite{14,15,16,17} and constituting the dark matter
\cite{3,6,7,18,19}, respectively.
Up to now, however, only more or less strong constraints on
the parameters of axions and axion-like particles have been
obtained. Specifically,
rather strong constraints on the coupling constants of
hadronic axions were found from astrophysics \cite{20}
(see Ref.~\cite{20a} for various models of hadronic axions).
In this context, one
could mention the constraints imposed by the neutrino data
of supernova SN~1987A \cite{21}, from stellar cooling
\cite{22,23},
from direct Chandra observation of the surface temperature of
isolated neutron stars \cite{24}, among others. It should be remembered,
however, that astrophysical data, such as the emission rate,
suffer from various uncertainties connected with dense matter
effects \cite{22}.

The model-independent constraints on the coupling constants of
neutrons to both hadronic and GUT axions were obtained from the
magnetometer measurements using spin-polarized K and ${}^3$He
atoms \cite{25}. In so doing, an exchange of one axion between
two neutrons has been used (see also Ref.~\cite{26} for other
constraints obtained from neutron physics).
Note that the effective interaction potential between two
fermions due to an exchange of one axion is spin-dependent,
which is common irrespective of whether the axion-fermion interaction
is described by the pseudo-vector or pseudo-scalar Lagrangian
\cite{27,28,29}. The model-independent laboratory constraints on
the coupling constants of GUT axion-like particles to nucleons
were obtained also
from E\"{o}tvos- and Cavendish-type experiments \cite{29,30,31},
and from measurements of the Casimir-Polder and Casimir forces
\cite{32,33,34,35} (see also Ref.~\cite{36}, for a review).
 Note also that a recent
Casimir-less experiment \cite{65}, where the Casimir force
is compensated in the measurement results, leads \cite{66,67,68a}
to stronger constraints in the same mass-range, which are
again valid for only the GUT axions.
In this case, the two-axion exchange between two nucleons has been
employed. The reason is that the test bodies used in the
E\"{o}tvos- and Cavendish-type experiments, and in measurements of
the Casimir interaction
are unpolarized. As a result, an additional interaction due to
one-axion exchange averages to zero. The effective potential due
to two-axion exchange between two fermions was derived
\cite{29,37,38}
with the assumption of a pseudo-scalar interaction Lagrangian
(we recall that the quantum field theory based on the
pseudo-vector Lagrangian
is nonrenormalizable).
This is the reason why all the constraints of
Refs.~\cite{29,30,31,32,33,34,35}
are applicable to the GUT axions (axion-like particles), but not
to the hadronic axions.

In this paper, we propose an experiment for measuring the effective
 Casimir pressure between two polarized conductive plates.
The suggested experiment would operate in much the same way as the
already performed ones with Au test bodies \cite{39,40,41,42} and
magnetic (Ni), but not magnetized, test bodies \cite{43,44,45}.
The experiments of Refs.~\cite{39,40,41,42} were performed by
 means of the micromechanical torsional oscillator consisting of
a Au-coated plate suspended at two opposite points on the midplane
by serpentine springs. Two independent electrodes located under
the plate were used to measure the capacitance between the plate
and electrodes and to induce oscillations in the plate at the
resonant frequency of the oscillator. Above the oscillator,
a large Au-coated sphere was mounted on the side of an optical
fiber in high vacuum. The Casimir force acting between a sphere
and a plate changes the resonant frequency of the oscillator,
and the measured frequency shift is proportional to the gradient
of the Casimir force between a sphere and a plate. In consequence
of the proximity force approximation \cite{48,49}, this frequency
shift is recalculated into the Casimir pressure between two plane
parallel plates (see Sec.~II). The experiments of
Refs.~\cite{43,44,45} were performed by means of the dynamic
atomic force microscope. In this case, a Au-coated sphere was
attached to a cantilever and oscillated harmonically above
a Au-coated plate in high vacuum. The shift of the resonant
frequency of an oscillator under the influence of the Casimir
force was again the immediately measured quantity. It was
recalculated into the gradient of the Casimir force between
a sphere and a plate or into
the Casimir pressure between two Au plates
(see the review \cite{49} for all experimental details).

First, we calculate an additional pressure which arises due to the
process of one-axion exchange between fermions belonging to
different plates.
In so doing, all possible directions of polarization of each plate
are considered. We show that, depending on the direction of
polarization, the additional pressure can be both repulsive and
attractive. Then we apply the obtained results to the case
of magnetic polarization and find constraints on an axion-electron
coupling constant which could be derived from an experiment of
this kind.
It is anticipated that the magnitude of the additional pressure due
to the one-axion exchange is smaller than the measurement error which
is assumed to be the same as in the experiments of
Refs.~\cite{40,41,44,45}. The obtained constraints on an axion-electron
coupling constant are much weaker than the  ones obtained from other
experiments. Thus, the Casimir experiment with magnetized test bodies is
not competitive.

An experiment for measuring the Casimir pressure between two
polarized plates
with aligned nuclear spins (i.e., possessing nuclear polarization)
 is shown to be more promising. We calculate the
 constraints
on an axion-neutron coupling constant, which could be obtained
by using the process of one-axion exchange from an experiment
with SiC plates and found them quite competitive in the
region of axion masses from $10^{-3}$ to 1\,eV.
These constraints will be equally valid for both the hadronic
and GUT axions (the constraints of Refs.~\cite{31,32,33,34}
obtained in this region of masses are applicable only to GUT
axions). Note that almost 100\% degree of polarization of
${}^{29}$Si nuclear spins in silicon carbide has been observed
recently \cite{46}. This makes the performance of
the proposed experiment quite
possible. We also calculate the constraints on the axion-neutron
coupling constant of the GUT axions which could be obtained
from measurements of the Casimir pressure between SiC plates with
aligned nuclear spins using simultaneously the processes of one-
and two-axion exchange.

The paper is organized as follows. In Sec.~II, we calculate the
pressure which arises between two polarized plates due to the
one-axion exchange between fermionic particles of their
materials. Section~III contains constraints on the axion-electron
coupling constant which could be obtained from measuring the
Casimir pressure between two magnetized plates. In Sec.~IV, we
propose an experiment for measuring the Casimir pressure between
SiC plates with aligned nuclear spins and calculate respective
constraints  on the axion-neutron coupling constants of both
hadronic and GUT axions. Section~V containts our conclusions and
discussion.

Throughout the paper we use the system of units where
$\hbar=c=1$.

\section{Calculation of the pressure between two polarized
plates due to one-axion exchange}

We consider two parallel material plates with densities $\rho_i$
($i=1,\,2$) and thicknesses $D_i$ separated
by a gap of width $a$. We assume that each plate contains a
fraction of atoms $\kappa_i$ polarized in some definite direction
(which can be different for different plates). The polarization
of atoms may originate from the spin polarization of either
electrons or nucleons (see Secs.~III and IV,
respectively).

Let us suppose
we have a fermion with spin $\mbox{\boldmath$\sigma$}_1/2$
(electron or nucleon) at a point $\mbox{\boldmath$r$}_1$ of the
first plate and a fermion with spin $\mbox{\boldmath$\sigma$}_2/2$
at a point $\mbox{\boldmath$r$}_2$ of the second plate.
For both the pseudo-vector and pseudo-scalar interactions of an
axion with fermions, the effective potential due to a one-axion
exchange between two fermions has the common form \cite{27,28,29}
\begin{eqnarray}
&&V(r_{12})=\frac{g^2}{16\pi m^2}\left[
(\mbox{\boldmath$\sigma$}_1\cdot\mbox{\boldmath$n$})
(\mbox{\boldmath$\sigma$}_2\cdot\mbox{\boldmath$n$})
\left(\frac{m_a^2}{r_{12}}+\frac{3m_a}{r_{12}^2}+
\frac{3}{r_{12}^3}\right)\right.
\nonumber \\
&&~~~~~~~~~~~~~
\left.
-(\mbox{\boldmath$\sigma$}_1\cdot\mbox{\boldmath$\sigma$}_2)
\left(\frac{m_a}{r_{12}^2}+\frac{1}{r_{12}^3}\right)\right]
e^{-m_ar_{12}}.
\label{eq1}
\end{eqnarray}
\noindent
Here, $g$ is the dimensionless interaction constant of an axion
with either an electron or nucleons (a neutron or a proton),
$m$ is the electron or nucleon mass,
 $r_{12}=|\mbox{\boldmath$r$}_1-\mbox{\boldmath$r$}_2|$ and
 $\mbox{\boldmath$n$}=(\mbox{\boldmath$r$}_1-\mbox{\boldmath$r$}_2)/r_{12}$.

We consider first one polarized atom of the first plate situated
at the point ($x_1,y_1,z_1$) above the upper surface of the second
 plate, which coincides with the coordinate plane ($x,y$).
 Let the spins of the polarized atoms of the second (lower) plate
be directed in the positive direction of the $z$ axis.
The polarized atoms of the second plate have the coordinates
($x_2,y_2,z_2$). Then, we have
\begin{eqnarray}
&&
r_{12}=\sqrt{(x_1-x_2)^2+(y_1-y_2)^2+(z_1-z_2)^2}
\nonumber \\
&&
~~~~
=
\sqrt{\rho^2+(z_1-z_2)^2},
\nonumber \\
&&
n_x=\frac{x_1-x_2}{r_{12}}=\frac{\rho\,\cos\varphi}{r_{12}},
\quad
n_z=\frac{z_1-z_2}{r_{12}},
\label{eq2}
\end{eqnarray}
\noindent
where we have introduced polar coordinates ($\rho,\varphi$)
in the coordinate plane ($x,y$) with the origin at the point
($x_1,y_1$).

In experiments for measuring the Casimir force the linear sizes of the
plates are much greater than
$a$, whereas the strongest constraints on the coupling
 constants of an axion are obtainable at the Compton wavelengths
 $m_a^{-1}\sim a$ \cite{32,33,34,35,36}.
Because of this, it is possible to consider plates as infinitely
large discs
and treat a polarized atom at the point ($x_1,y_1,z_1$) as situated
above its center.
Then, the interaction potential between an atom and a second plate due
to the one-axion exchange is proportional to the fraction
$\kappa_2$ of the polarized atoms of the second plate and
takes the form
\begin{equation}
V_1(z_1)=\kappa_2\frac{\rho_2}{m_{\rm H}}\int_{-D_2}^{0}
dz_2\int_{0}^{2\pi}d\varphi\int_{0}^{\infty}\rho d\rho
V(\rho,z_1,z_2),
\label{eq3}
\end{equation}
\noindent
where $m_{\rm H}$ is the mass of an atom of hydrogen, so that
$\rho_i/m_{\rm H}$ is the number of atoms in the unit volume of
the plate $i$ and
$V$ is defined in Eq.~(\ref{eq1}).

Now we assume that the polarization of an atom of the first plate
is directed either along the $z$ axis or in opposition to it.
Then, one obtains
\begin{equation}
(\mbox{\boldmath$\sigma$}_i\cdot\mbox{\boldmath$n$})=
\sigma_i\frac{z_1-z_2}{r_{12}}, \quad
(\mbox{\boldmath$\sigma$}_1\cdot\mbox{\boldmath$\sigma$}_2)=
\sigma_1\sigma_2
\label{eq4}
\end{equation}
\noindent
or
\begin{eqnarray}
&&
(\mbox{\boldmath$\sigma$}_1\cdot\mbox{\boldmath$n$})=
-\sigma_1\frac{z_1-z_2}{r_{12}}, \quad
(\mbox{\boldmath$\sigma$}_2\cdot\mbox{\boldmath$n$})=
\sigma_2\frac{z_1-z_2}{r_{12}},
\nonumber \\
&&
(\mbox{\boldmath$\sigma$}_1\cdot\mbox{\boldmath$\sigma$}_2)=
-\sigma_1\sigma_2,
\label{eq5}
\end{eqnarray}
\noindent
respectively.

Substituting Eqs.~(\ref{eq1}), (\ref{eq4}) and (\ref{eq5}) in
Eq.~(\ref{eq3}), and integrating with respect to $\varphi$,
for the two polarization directions of the atom one obtains
\begin{eqnarray}
&&
V_1(z_1)=\pm\sigma_1\sigma_2\frac{\kappa_2\rho_2}{m_{\rm H}}
\frac{g^2}{8m^2}\int_{-D_2}^{0}\!\!\!dz_2
\int_{0}^{\infty}\!\!\!\!\rho d\rho
e^{-m_ar_{12}}
\label{eq6} \\
&&
~~~\times\left[
\frac{m_a^2(z_1-z_2)^2}{r_{12}^3}+
\frac{3m_a(z_1-z_2)^2}{r_{12}^4}+
\frac{3(z_1-z_2)^2}{r_{12}^5}-
\frac{m_a}{r_{12}^2}-\frac{1}{r_{12}^3}\right],
\nonumber
\end{eqnarray}
\noindent
where $r_{12}$ is given in Eq.~(\ref{eq2}).

It is convenient to rewrite Eq.~(\ref{eq6}) in terms of the new
variable $u=m_ar_{12}$, which varies from $u_1=m_a(z_1-z_2)$ to
$\infty$:
\begin{equation}
V_1(z_1)=\pm g^2\frac{\sigma_1\sigma_2\kappa_2\rho_2m_a}{8m^2m_{\rm H}}
\int_{-D_2}^{0}\!\!\!dz_2I(u_1),
\label{eq7}
\end{equation}
\noindent
where
\begin{equation}
I(u_1)\equiv\int_{u_1}^{\infty}\!\!du e^{-u}\left(
-\frac{1}{u}+\frac{u_1^2-1}{u^2}+\frac{3u_1^2}{u^3}
+\frac{3u_1^2}{u^4}\right).
\label{eq8}
\end{equation}
\noindent
All the integrals in Eq.~(\ref{eq8}) are simply calculated
\cite{47} with the result
\begin{equation}
I(u_1)=e^{-u_1}=e^{-m_a(z_1-z_2)}.
\label{eq9}
\end{equation}
\noindent
Substituting Eq.~(\ref{eq9}) in Eq.~(\ref{eq7}), one arrives at
\begin{eqnarray}
&&
V_1(z_1)=\pm g^2\frac{\sigma_1\sigma_2\kappa_2\rho_2m_a}{8m^2m_{\rm H}}
\int_{-D_2}^{0}\!\!\!dz_2e^{-m_a(z_1-z_2)}
\nonumber \\
&&
~~~~~~
=\pm g^2\frac{\sigma_1\sigma_2\kappa_2\rho_2}{8m^2m_{\rm H}}
e^{-m_az_1}(1-e^{-m_aD_2}).
\label{eq10}
\end{eqnarray}

Note that if an atom of the first plate would be polarized not
perpendicular, but along the surface of the second plate
(for instance, along the $x$ axis), this results in
\begin{equation}
(\mbox{\boldmath$\sigma$}_1\cdot\mbox{\boldmath$\sigma$}_2)=0,
\quad
(\mbox{\boldmath$\sigma$}_1\cdot\mbox{\boldmath$n$})
(\mbox{\boldmath$\sigma$}_2\cdot\mbox{\boldmath$n$})=
\sigma_1\sigma_2\frac{\rho\cos\varphi(z_1-z_2)}{r_{12}^2}.
\label{eq11}
\end{equation}
\noindent
Then, both terms in Eq.~(\ref{eq1}) do not contribute to the
potential (\ref{eq3}) (the first one turns into zero after an
integration in $\varphi$, and the second one is identically equal
to zero). Thus, there is no atom-plate interaction due to the
exchange of one axion in this case.

Now we consider some other possible cases which could lead to
the interaction between a polarized atom of the first plate and the
second plate. Let the spins of the polarized atoms of the second
(lower) plate be directed along its surface
(for instance, along the $x$ axis). The nontrivial situations
are when the spin of a polarized atom of the first plate is either
parallel or antiparallel to it. In these cases, we have
\begin{equation}
(\mbox{\boldmath$\sigma$}_1\cdot\mbox{\boldmath$n$})
(\mbox{\boldmath$\sigma$}_2\cdot\mbox{\boldmath$n$})=
\sigma_1\sigma_2\frac{\rho^2\cos^2\varphi}{r_{12}^2}
\label{eq12}
\end{equation}
\noindent
or
\begin{equation}
(\mbox{\boldmath$\sigma$}_1\cdot\mbox{\boldmath$n$})
(\mbox{\boldmath$\sigma$}_2\cdot\mbox{\boldmath$n$})=
-\sigma_1\sigma_2\frac{\rho^2\cos^2\varphi}{r_{12}^2},
\label{eq13}
\end{equation}
\noindent
respectively, and the same respective results, as in
Eqs.~(\ref{eq4}) and (\ref{eq5}), are valid for
$(\mbox{\boldmath$\sigma$}_1\cdot\mbox{\boldmath$\sigma$}_2)$.

Substituting Eqs.~(\ref{eq1}), (\ref{eq12}) and (\ref{eq13}) in
Eq.~(\ref{eq3}),
for the two directions of polarization of an atom belonging to the
first plate we find
\begin{eqnarray}
&&
V_1(z_1)=\pm\sigma_1\sigma_2\frac{\kappa_2\rho_2}{m_{\rm H}}
\frac{g^2}{16\pi m^2}\int_{-D_2}^{0}\!\!\!dz_2
\int_{0}^{2\pi}\!\!\!d\varphi
\int_{0}^{\infty}\!\!\!\!\rho d\rho
e^{-m_ar_{12}}
\nonumber \\
&&
~~~\times\left[
\frac{m_a^2\rho^2\cos^2\varphi}{r_{12}^3}+
\frac{3m_a\rho^2\cos^2\varphi}{r_{12}^4}\right.
\nonumber \\
&&
~~~~~~~~~~~~~~\left.+
\frac{3\rho^2\cos^2\varphi}{r_{12}^5}-
\frac{m_a}{r_{12}^2}-\frac{1}{r_{12}^3}\right].
\label{eq14}
\end{eqnarray}
\noindent
Carrying out the
integration in Eq.~(\ref{eq14}) with respect to $\varphi$, and
taking into account the definition of $r_{12}$ in
Eq.~(\ref{eq12}), one obtains
\begin{eqnarray}
&&
V_1(z_1)=\pm\sigma_1\sigma_2\frac{\kappa_2\rho_2}{m_{\rm H}}
\frac{g^2}{16 m^2}\int_{-D_2}^{0}\!\!\!dz_2
\int_{0}^{\infty}\!\frac{\rho d\rho}{r_{12}}
e^{-m_ar_{12}}
\nonumber \\
&&
~~~\times\left[m_a^2+\frac{m_a}{r_{12}}+
\frac{1-m_a^2(z_1-z_2)^2}{r_{12}^2}\right.
\nonumber \\
&&
~~~~~~~~~~~~~\left.-
\frac{3m_a(z_1-z_2)^2}{r_{12}^3}-
\frac{3(z_1-z_2)^2}{r_{12}^4}\right].
\label{eq15}
\end{eqnarray}

Now we introduce in Eq.~(\ref{eq15}) the new integration variable
$u=m_ar_{12}$ [the same as in Eq.~(\ref{eq7})] and obtain
\begin{equation}
V_1(z_1)=\pm g^2\frac{\sigma_1\sigma_2\kappa_2\rho_2m_a}{16 m^2m_{\rm H}}
\int_{-D_2}^{0}\!\!\!dz_2\tilde{I}(u_1),
\label{eq16}
\end{equation}
\noindent
where
\begin{equation}
\tilde{I}(u_1)\equiv\int_{u_1}^{\infty}\!\!du e^{-u}\left(
1+\frac{1}{u}+\frac{1-u_1^2}{u^2}-\frac{3u_1^2}{u^3}
-\frac{3u_1^2}{u^4}\right).
\label{eq17}
\end{equation}
\noindent
Calculation of all integrals here \cite{47} leads to the result
$\tilde{I}(u_1)=0$, and, thus, $V_1(z_1)=0$, as well.
One can conclude that if the polarizations of an atom and a
plate are parallel to the plate surface there is no atom-plate
force due to the process of one-axion exchange.

One more case to consider is when atoms of the second plate are
polarized along its surface (for instance, along the $x$ axis),
whereas the atom of the first plate is polarized in the perpendicular
direction (along the $z$ axis). In this case, we obtain the same
results as in Eq.~(\ref{eq11}), leading to the zero interaction
potential (\ref{eq3}).

Thus, we have considered six different options and found that
nonzero interaction potentials (\ref{eq10}) of opposite sign
arise when the atomic polarizations are perpendicular to the plate
surface and directed either in one direction or in the opposite
directions.
If the direction of an atomic polarization makes some arbitrary angle
to the plate surface, only its component perpendicular to the
surface contributes to the interaction potential due to a
one-axion exchange.

{}From Eq.~(\ref{eq10}) it is easy to obtain the interaction
energy per unit area
of two parallel plates due to a one-axion exchange between the
polarized atoms belonging to these plates. For this purpose, we
integrate Eq.~(\ref{eq10}) over the volume
of the first plate with the
coefficient taking into account the number of polarized atoms
per unit volume. The result is
\begin{eqnarray}
&&
E(a)=\pm g^2\frac{\sigma_1\sigma_2}{8m^2}
\frac{\kappa_1\kappa_2\rho_1\rho_2}{m_{\rm H}^2}
(1-e^{-m_aD_2})
\int_{a}^{D_1+a}\!\!\!dz_1e^{-m_az_1}
\nonumber \\
&&
~~~~~
=\pm g^2\frac{\sigma_1\sigma_2}{8m^2m_a}
\frac{\kappa_1\kappa_2\rho_1\rho_2}{m_{\rm H}^2}
e^{-m_aa}
(1-e^{-m_aD_1})(1-e^{-m_aD_2}).
\label{eq18}
\end{eqnarray}

The most precise experiments on measuring the Casimir
interaction \cite{39,40,41,42,43,44,45} exploit the
configuration of a sphere above a plate rather than of two
parallel plates. In so doing, the immediately measured
quantity is not the force $F_{sp}$ acting between a sphere
and a plate, but the force gradient $\partial F_{sp}/\partial a$.
Using the proximity force approximation, which is very accurate
under the condition $a\ll R$ for both the Casimir and
Yukawa-type forces \cite{48,49,50,51}, one obtains
\begin{equation}
\frac{\partial F_{sp}(a)}{\partial a}=
2\pi R\frac{\partial E(a)}{\partial a}=
-2\pi RP(a).
\label{eq19}
\end{equation}
\noindent
This equation expresses the force gradient between a sphere and
a plate via the pressure $P$ in the
configuration of two parallel plates.
Then the results of experiments \cite{39,40,41,42,43,44,45}
can be interpreted as measurements of the effective Casimir
pressure between two parallel plates.

{}From Eqs.~(\ref{eq18}) and (\ref{eq19}), the additional pressure
between two parallel plates due to a one-axion exchange is
found to be
\begin{equation}
P^{(1)}(a)=
=\mp g^2\frac{\sigma_1\sigma_2}{8m^2}
\frac{\kappa_1\kappa_2\rho_1\rho_2}{m_{\rm H}^2}
e^{-m_aa}
(1-e^{-m_aD_1})(1-e^{-m_aD_2}).
\label{eq20}
\end{equation}
\noindent
This result is used in Secs.~III and IV for the cases of
magnetized plates and
plates with aligned nuclear spins.

\section{Constraining axion-electron coupling constant from
Casimir experiment with magnetized test bodies}

In the experiment of Refs.~\cite{44,45}, the gradient of the
Casimir force has been measured between the surfaces of a
sphere and a plate coated by sufficiently thick layers of
magnetic metal Ni. The measurements were performed by means of
an atomic force microscope. The measurement data were found
to be in excellent agreement with the theoretical predictions of the
Lifshitz theory using the tabulated optical data of Ni \cite{52}
extrapolated to lower frequencies by means of the plasma model.
The theoretical prediction using the Drude model for an
extrapolation were excluded by the measurement data. Recently,
these
 results have been conclusively confirmed in Ref.~\cite{53}
using a configuration, where theoretical predictions computed
by means of the plasma and Drude models differ by up to a factor
of 1000 \cite{54}.

The test bodies used in Refs.~\cite{44,45} are magnetic, but not
magnetized. What is more, it was shown \cite{44,45} that even for
fully magnetized test bodies the gradient of the magnetic force
acting between them is much below the instrumental sensitivity.
This is because the magnetization results in a spatially
homogeneous magnetic force at the submicrometer separations.
This force leads to very minor contributions to the measured
force gradient. Thus, it is possible to repeat the experiment
\cite{44,45} with no other changes, but with the magnetized
Ni test bodies in perpendicular direction to their surfaces.

For the magnetized Ni films, the magnetic moment of each atom
is determined by a single electron of mass $m=m_e$.
Then, the interaction constant $g$ in Eq.~(20) has the meaning of
an axion-electron coupling constant $g_{ae}$.
Note that for free Ni atoms $\sigma_1=\sigma_2=\sigma=1$,
whereas for Ni atoms included in a crystal lattice (as in our
case) $\sigma=0.6$ \cite{55}. Taking into account
also that $\kappa_1=\kappa_2=1$, Eq.~(\ref{eq20}) takes
the form
\begin{equation}
|P^{(1)}(a)|=
\frac{1}{8} g_{ae}^2\left(\frac{\sigma}{m_e}\right)^2
\left(\frac{\rho_{\rm Ni}}{m_{\rm H}}\right)^2
e^{-m_aa}
(1-e^{-m_aD_1})(1-e^{-m_aD_2}),
\label{eq21}
\end{equation}
\noindent
where $\rho_1=\rho_2=\rho_{\rm Ni}=8.9\times 10^3\mbox{kg/m}^3$.
The thicknesses of Ni films used in the experiment of
Refs.~\cite{44,45} are the following:
$D_1=250\,$nm and $D_1=210\,$nm.

The constraints on $g_{ae}$ could be obtained from the assumption
that in the experiment for measuring the gradient of the Casimir
force between magnetized Ni test bodies the experimental data
are in agreement with the same theory,
as in Refs.~\cite{44,45}. This means that
no any additional force is detected, i.e., the magnitude of
the pressure (\ref{eq21}), due to one-axion exchange, satisfies the
condition
\begin{equation}
|P^{(1)}(a)|<\frac{\Delta^{\!\rm tot}F_C^{\prime}}{2\pi R}=
\Delta^{\!\rm tot}P_C,
\label{eq22}
\end{equation}
\noindent
where $\Delta^{\!\rm tot}F_C^{\prime}$ and $\Delta^{\!\rm tot}P_C$
 are the total experimental errors in the measurements of the
gradient of the Casimir force and of the effective Casimir
pressure, respectively. In Refs.~\cite{44,45}
$\Delta^{\!\rm tot}F_C^{\prime}=1.2\,\mu$N/m at all separations
considered, which results in  $\Delta^{\!\rm tot}P_C=3.1\,$mPa.

Numerical analysis of Eqs.~(\ref{eq21}) and (\ref{eq22}) shows
that the strongest constraints on the quantity $g_{ae}^2/(4\pi)$
 are obtainable in the region of axion masses from 0.6 to 2\,eV.
Thus, for $m_a=0.65$, 0.8, 1, 1.3, and 2\,eV, $g_{ae}^2/(4\pi)$
should be smaller than $1.0\times 10^{-9}$, $9.5\times 10^{-10}$,
$9.2\times 10^{-10}$, $1.0\times 10^{-9}$, and $1.6\times 10^{-9}$,
respectively. For comparison, the constraints on $g_{ae}$ obtained
 in Ref.~\cite{56} for solar axions produced by the Compton
 process and
bremsstrahlung are given by $m_ag_{ae}\leq 3.1\times 10^{-7}\,$eV.
 For the axion mass $m_a=1\,$eV, this results in
$g_{ae}^2/(4\pi)\leq 7.6\times 10^{-15}$, which is a much stronger
constraint than the one following from Eq.~(\ref{eq22}).
We conclude that experiments for measuring the Casimir interaction
between two magnetized Ni test bodies are not a promising
method for obtaining
stronger constraints on the axion-electron coupling constant.

\section{Proposed experiment using test bodies with aligned
{\protect \\} nuclear spins}

Another possibility for obtaining constraints on the coupling
constants of an axion from the Casimir effect using the process
of one-axion exchange is to employ test bodies with nuclear
polarization. It is well known that spin polarization can be
transferred from electrons to nuclei. For metals this effect was
first demonstrated theoretically and experimentally in the
classical papers \cite{57,58}. In succeeding years, many different
 techniques for the dynamic nuclear polarization have been
 suggested \cite{59,60,61}. It has been made possible to achieve
 high degrees of nuclear polarization, even at room temperature.
 This allowed producing nuclear polarized targets for particle
 physics and many other important applications.

Here, we propose a measurement of the effective Casimir pressure
 between two parallel plates made of silicon carbide (SiC) with
 aligned nuclear spins of Si. On the one hand, it was recently
 shown \cite{46} that optically pumped nuclear polarization of
 ${}^{29}$Si nuclear spins in SiC can achieve 99\% at room
 temperature. On the other hand, SiC is a semiconductor which
 can be doped both $n$-type and $p$-type \cite{62}.
Recently, it was used as a plate material in measurements of the
Casimir force \cite{63}.

\subsection{Constraints on axion-neutron coupling constant using
the process of one-axion exchange}

Here, we propose an experiment for measuring the effective
Casimir pressure between two all-silicon-carbide plates with
aligned nuclear spins. Note that the nuclear spin of
${}^{29}$Si is equal to 1/2 ($\sigma_1=\sigma_2=1$) due to the
presence of one neutron with an uncompensated spin.
The natural abundance of ${}^{29}$Si is 4.6832\%.
There are, however, nanotechnology methods for growing
isotopically controlled bulk Si \cite{64}. Because of this,
below we obtain the prospective constraints on an axion-neutron
coupling constant $g_{an}$, using various values of
$\kappa_1=\kappa_2=\kappa$. Taking into account that the
actual thicknesses of the plates $D_1$ and $D_2$ are not yet
available, we put them equal to infinity, i.e., consider two
semispaces. This assumption works well as long as the
Compton wavelength of an axion is much less than $D$.
It is easy to take into consideration the actual values of $D_i$,
as well as the boundary effects arising due to the finitness
of the plate area \cite{34}.

Taking into account all the above considerations,
Eq.~(\ref{eq21}) for the
magnitude of effective pressure due to one-axion exchange can be
written in the form
\begin{equation}
|P^{(1)}(a)|=
=\frac{1}{8} g_{an}^2
\left(\frac{\kappa\rho_{\rm SiC}}{m_nm_{\rm H}}\right)^2
e^{-m_aa},
\label{eq23}
\end{equation}
\noindent
where the density of SiC is $\rho_{\rm SiC}=3.21\mbox{g/cm}^3$.

An experiment for measuring the effective Casimir pressure
between nuclear polarized SiC plates could be performed similarly
to Refs.~\cite{39,40,41}, using a micromechanical torsional
oscillator (this tecnique leads to a more precise results than
that using
an atomic force microscope). The total experimental error in
Refs.~\cite{40,41} is separation-dependent. The strongest
constraints on $g_{an}$ are obtainable at the separation distance
$a=300\,$nm, where $\Delta^{\!\rm tot}P_C=0.22\,$mPa \cite{40,41}.
Then, the constraints on the quantity $g_{an}^2/(4\pi)$ were
found numerically from Eqs.~(\ref{eq22}) and (\ref{eq23}).

In Fig.~1, the obtained constraints are shown as functions of the
axion mass by the four dashed lines computed from top to bottom
for the fractions $\kappa$ of ${}^{29}$Si equal to 0.046832
(the natural abundance), 0.1, 0.5, and 1.0, respectively.
The regions of [$m_a,g_{an}^2/(4\pi)$] plane above each line
are excluded by the experimental results of the proposed
experiment,
and the regions below each line are allowed.
As is seen in Fig.~1, the constraints become stronger with
increasing fraction  of ${}^{29}$Si isotope atoms whose
nuclear spins are aligned. The strongest constraint shown
by the bottom dashed line  is
$g_{an}^2/(4\pi)\leq 4.43\times 10^{-5}$.
It is valid for axions with mass $m_a=0.0126\,$eV.

The constraints following from this experiment would be valid
for both hadronic and GUT axions.
For comparison purposes, the solid line~1 in Fig.~1 shows the
constraints on an axion-neutron coupling constant obtained
\cite{25} by means of a magnetometer, using spin polarized
K and ${}^3$He atoms. These constraints found in the wide region
of axion masses from $10^{-10}$ to $6\times 10^{-6}\,$eV are
also valid for both the hadronic and GUT axions.
Similarly to our constraints, they were
derived using the effective potential (\ref{eq1}).
As can be seen in Fig.~1, the constraints obtainable from
the proposed measurements of the Casimir interaction between
the test bodies with aligned nuclear spin and the constraints
obtained from the magnetometer measurements complement each
other nicely.

In Fig.~1 we also plot the constraints valid for only the
GUT axions obtained from the gravitational experiments of
Cavendish type \cite{30,31} (the solid line 2) and experiments
on measuring the Casimir pressure between unpolarized test
bodies \cite{34} (the solid line 3) using the process of
two-axion exchange. As is seen in Fig.~1, the proposed
experiment could strengthen the previously known constraints
on $g_{an}$ for GUT axions in the region of axion masses
$4.4\,\mbox{meV}<m_a<10\,$eV and to extand them to both
hadronic and GUT axions.

\subsection{Constraints on the coupling constant of
axion-like particles to neutrons using
the processes on one- and two-axion exchange}

Now we calculate the constraints on $g_{an}$ obtainable from
the proposed Casimir experiment with nuclear polarized plates
if the processes on one- and two-axion exchange are taken
into account. The inclusion of the two-axion exchange into
consideration restricts the region of applicability of the obtained
results to the case of GUT axions only (see Sec.~I).
Although the interaction potential due to two-axion
exchange is weaker than for an exchange of one axion, all
nucleons of the test bodies contribute to it, and not just a
small fraction of them with aligned spins.

The effective potential due to two-axion exchange between two
nucleons (either protons or neutrons) belonging to the first
and second plates is given by \cite{29,37,38}
\begin{equation}
V_{kl}(r_{12})=-\frac{g_{ak}^2g_{al}^2m_a}{32\pi^3 m^2}
\frac{K_1(2m_ar_{12})}{r_{12}^2},
\label{eq24}
\end{equation}
\noindent
where $g_{ak}$ and $g_{al}$ are the coupling constants of an
axion-like particle to a proton ($k,l=p$) or a neutron
($k,l=n$), $m=(m_p+m_n)/2$ is the mean nucleon mass, and
$K_1(z)$ is the modified Bessel function of the second kind.

An integration of Eq.~(\ref{eq24}) over the volumes of both
 plates (semispaces) leads to the
 following effective pressure due to
the process of  two-axion exchange \cite{34}:
\begin{equation}
P^{(2)}(a)=-\frac{C^2}{2m^2m_{\rm H}^2}
\int_{1}^{\infty}\!\!\!du\frac{\sqrt{u^2-1}}{u^2}
e^{-2m_aau}.
\label{eq25}
\end{equation}
\noindent
Here, the coefficient $C$ for the plates made of SiC is
defined as
\begin{equation}
C=\rho_{\rm SiC}\left(\frac{g_{ap}^2}{4\pi}
\frac{Z}{\mu}+\frac{g_{an}^2}{4\pi}
\frac{N}{\mu}\right),
\label{eq26}
\end{equation}
\noindent
where $Z$ and $N$ are the numbers of protons and neutrons in the
molecule SiC and $\mu=m_{\rm SiC}/m_{\rm H}$, with $m_{\rm SiC}$
being the mass of the SiC molecule. The values of
$Z/\mu$ and $N/\mu$ for the first 92 elements of the Periodic
Table (as well as the algorithms for calculating these
quantities for molecules) are contained in Ref.~\cite{68}.

We note that Eq.~(\ref{eq25}) depends on two unknown interaction
 constants, $g_{an}$ and $g_{ap}$.
As shown in Refs.~\cite{32,33,34,35}, the weakest constraints on
$g_{an}$ are obtained under the condition $g_{ap}\ll g_{an}$.
Being conservative, we use this condition now, which leads from
Eq.~(\ref{eq26}) to the result
\begin{equation}
C\approx \rho_{\rm SiC}\frac{g_{an}^2}{4\pi}
\frac{N}{\mu},
\label{eq27}
\end{equation}
\noindent
where for SiC one finds $N=20.11987$ and $\mu=39.78539$ \cite{68}.

The constraints on the coupling constant $g_{an}$ of GUT axions
(axion-like particles) can now be obtained from the equation
\begin{equation}
|P^{(1)}(a)|+|P^{(2)}(a)|\leq \Delta^{\!\rm tot}P_C(a),
\label{eq28}
\end{equation}
\noindent
where $P^{(1)}$ and $P^{(2)}$ are defined in Eqs.~(\ref{eq23}) and
(\ref{eq25}). In so doing, we assume that the first and second
plates are polarized in the opposite directions, so that
$P^{(1)}$ and $P^{(2)}$ are negative, which corresponds to an
attraction. The strongest constraints on $g_{an}$ are again
obtainable at $a=300\,$nm, where
$\Delta^{\!\rm tot}P_C(a)=0.22\,$mPa \cite{40,41}.

In Fig.~2 we plot the obtained prospective constraints on
$g_{an}$ representing them
by the dashed lines as functions of the axion mass.
The top dashed line is computed for the fraction of polarized
atoms $\kappa$ between 0 and 0.1. In this case, the computational
results do not depend on whether the plates are polarized or
unpolarized. The reason is that the total axionic pressure is
determined by the two-axion exchange. The intermediate and
bottom dashed lines are computed for the cases $\kappa=0.5$
and 1, respectively. Here, the nuclear polarization contributes
essentially to a result which is determined by the joint
action of one- and two-axion exchange.

For comparison purposes, the solid line in Fig.~2 reproduces
the constraints on $g_{an}\gg g_{ap}$ obtained \cite{34} from
an experiment on measuring the effective Casimir pressure between
Au plates, using the process of two-axion exchange.
As can be seen in Fig.~2, the intermediate dashed line
($\kappa=0.5$)
shows the constraints of almost the same strength.
In this case, the advantage of using the processes of both one-
and two-axion exchange is reduced to zero due to the lower density
of SiC, as compared with Au. However, the bottom dashed line
($\kappa=1$) shows up to a factor of 3.7 stronger
constraints in comparison with the solid line. Thus, the
proposed experiment allows not only to extend the previously
known constraints to the case of hadronic axions, but also
to strengthen some results obtained previously for the GUT axions.

\section{Conclusions and discussion}

In the foregoing, we have found the additional pressure which
arises between two polarized parallel plates due to the process
of one-axion exchange between the constituent fermions.
Only the components of the polarizations which are
normal to the plates  are
shown to contribute to the pressure. If the polarizations are
unidirectional, the pressure is repulsive.
If the polarizations are directed in opposite directions, then
the pressure has  the same
magnitude, but it is attractive.

The obtained results were applied to the cases of
magnetized plates and
 plates with aligned nuclear spins. In the first case, the
performance of the experiment for measuring the effective Casimir
pressure between two magnetized plates would constrain the
axion-electron coupling constant $g_{ae}$. We have calculated
the constraints on $g_{ae}$ obtainable in this way by using the
parameters of similar experiment already performed and found
them not enough competitive. In the second case, the
measurement of
the effective Casimir pressure between two plates possessing
nuclear polarization would constrain the axion-nucleon
coupling constants $g_{an}$ and $g_{ap}$.

We have proposed an experiment for measuring the effective Casimir
pressure between SiC plates which have already been successfully
used in Casimir experiments \cite{63}. This material was
preferred because it allows almost 100\% nuclear polarization of
${}^{29}$Si nuclei \cite{46} due to the presence of one neutron
with an uncompensated spin. The respective constraints on an
axion-neutron coupling constant $g_{an}$, obtainable for both
hadronic and GUT axions from the suggested experiment, were
calculated for various fractions of ${}^{29}$Si using typical
already obtained experimental parameters. These
constraints are shown to be
of the same strength, as obtained previously for
GUT axions only using the process of two-axion exchange from
experiments on measuring the effective Casimir pressure between
two Au plates. The discussed constraints would be complementary to the
constrainsts, obtained from the magnetometer measurements for
 axions of lower masses,  which are also valid for both hadronic
and GUT axions.

Finally, we have calculated the constraints on the axion-neutron
coupling constant $g_{an}$ of GUT axions following from the
proposed experiment with SIC plates if both processes of one-
and two-axion exchange are taken into account. It is shown that,
under some conditions, constraints stronger than those found
previously from the experiment with two Au plate are obtainable.

In Secs.~III and IV we have considered two different materials
(Ni and SiC) possessing the electron and nuclear polarizations,
respectively. If some material possesses both kinds of
polarizations, this does not reduce a sensitivity of the proposed
experiment to the axion-neutron interaction. This conclusion
remains valid even if the electron and nuclear polarizations cancel
each other. The point is that interactions of axions with the
magnetic moments of electrons and neutrons are quite independent.
Furthermore, at the experimental separations below $1\,\mu$m,
the magnetic field, if any, is space homogeneous and does not
contribute to the measured force gradient. One can also consider
not the case of nuclear spins polarized in the same directions
(as we assumed above),
but a periodic arrangement with zero net polarization in both
plates. In this case a nonzero force due to one-axion exchange
between nucleons can arise only under a condition that the
periods in both plates are equal. Then, depending on a phase
shift between the periodic structures in both plates, the
resulting pressure due to one-axion exchange would vary between
some $-P_{\max}$ and $P_{\max}$. The value of $P_{\max}$ can be
calculated with the help of Fourier expansions similar to
Ref.~\cite{45}. The experimental procedure using the periodically
arranged nuclear spins would demand the lateral scanning of the
first test body relative to the second one to find out whether
or not the measured pressure depends on a phase shift.

To conclude, experiments for measuring the Casimir interaction
between test bodies made of different materials remain to be
prospective for constraining the predictions of fundamental
physical theories. In the past, strong constraints on the
Yukawa-type corrections to Newtonian gravity have been obtained
in this way (see, e.g., Refs.~\cite{48,69,70,71}).
At the present time, measurements of the Casimir interaction are
helpful for constraining the coupling constants of axions and
axion-like particles. One might expect that new experiments for
measuring the Casimir force will lead to further progress in
both these directions.

\section*{Acknowledgments}

The authors of this work acknowledge CNPq (Brazil) for
 partial financial support (the Grants 307596/2015--0 and
 308150/2015--5).
G.L.K.\ and V.M.M.\ are grateful to the Department
of Physics of the Federal University of
Para\'{\i}ba (Jo\~{a}o Pessoa, Brazil) for kind hospitality.


\newpage
\begin{figure}[b]
\vspace*{-8cm}
\centerline{\hspace*{2cm}
\includegraphics{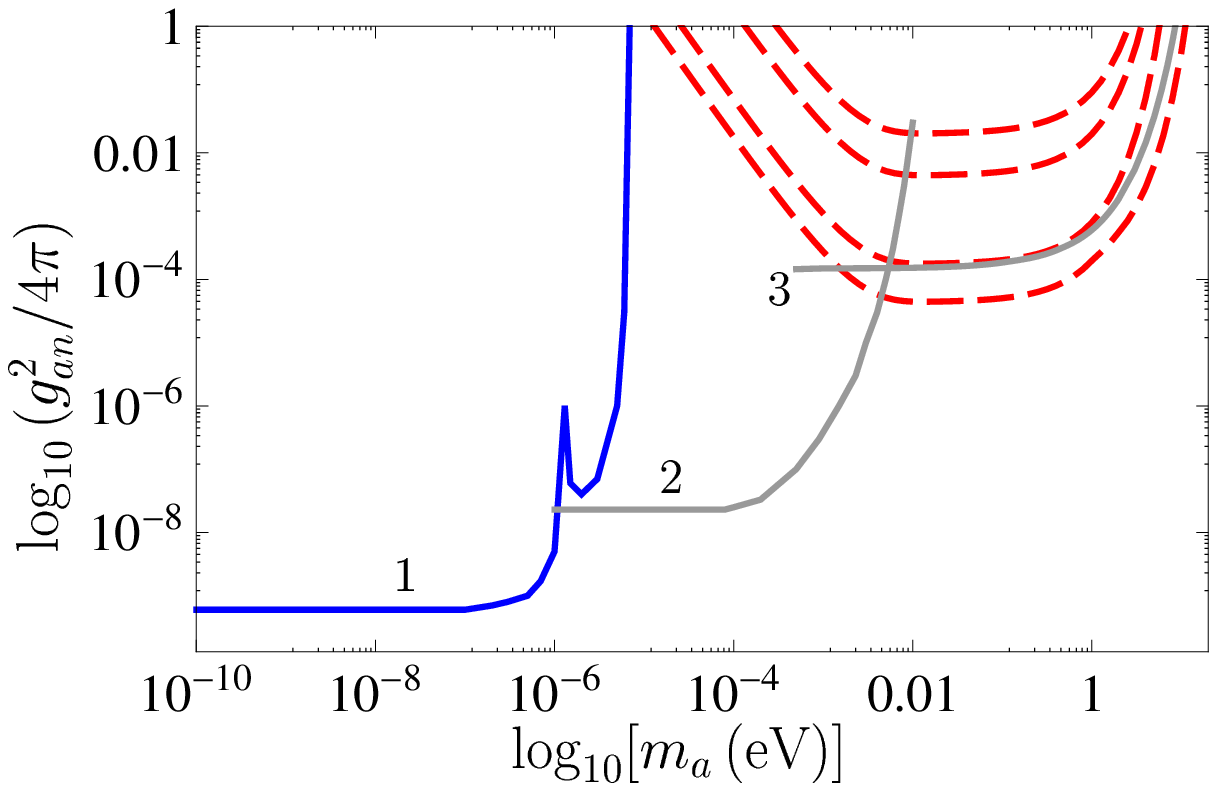}
}
\vspace*{-9cm}
\caption{(Color online)
Constraints on the  axion-neutron coupling constant
of both hadronic and GUT axions obtained \cite{25} from the
magnetometer measurements (the solid line 1) and
obtainable from the proposed experiment for measuring
the effective Casimir pressure between SiC plates with
aligned nuclear spins (the dashed lines from top to
bottom correspond to the fractions of ${}^{29}$Si
isotope equal to 0.046832, 0.1, 0.5, and 1.0,
respectively) using the process of one-axion exchange
 are shown as functions of the axion mass.
The solid lines 2 and 3 show the constraints for only
the GUT axions found from the Cavendish-type experiment
\cite{30,31} and from measurements of the Casimir
pressure \cite{34}, respectively, using the process of
two-axion exchange.
 The regions  of the plane above each line
are excluded and below each line they are allowed.
}
\end{figure}
\begin{figure}[b]
\vspace*{-8cm}
\centerline{\hspace*{2cm}
\includegraphics{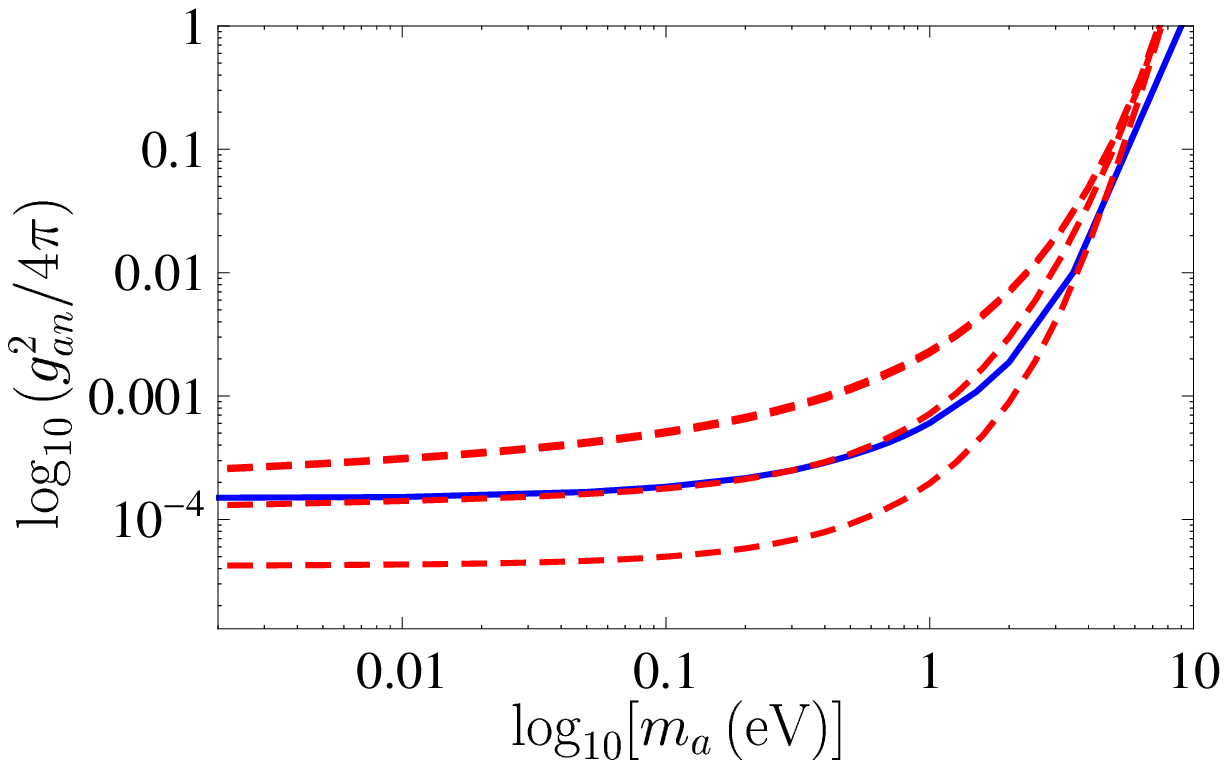}
}
\vspace*{-9cm}
\caption{(Color online)
Constraints on the axion-neutron coupling constant
of GUT axions already
obtained \cite{34} from measuring the effective
Casimir pressure between two Au plates (the solid line)
using the process of two-axion exchange and obtainable
from similar proposed experiment using SiC plates with
aligned nuclear spins (the dashed lines from top to
bottom correspond to the fractions of ${}^{29}$Si
isotope  $\leq0.1$, 0.5, and 1.0,
respectively) using the processes of one- and
two-axion exchange
 are shown as functions of the axion mass.
 The regions  of the plane above each line
are excluded and below each line they are allowed.
}
\end{figure}
\end{document}